\documentclass[12pt]{article}
\usepackage{graphicx}
\usepackage{amsmath}
\usepackage{amssymb}
\usepackage{pdfpages}

\usepackage{scicite}

\usepackage{times}

\newenvironment{sciabstract}{%
\begin{quote} \bf}
{\end{quote}}

\title{Non-adiabatic exciton-phonon coupling in Raman spectroscopy of layered materials}

\author
{Sven Reichardt,$^{1\ast}$ Ludger Wirtz$^{1}$\\
\\
\normalsize{$^{1}$Department of Physics and Materials Science, University of Luxembourg}\\
\normalsize{1511 Luxembourg, Luxembourg}\\
\\
\normalsize{$^\ast$To whom correspondence should be addressed; E-mail:  sven.reichardt@uni.lu}
}

\date{}

\begin{document} 


\baselineskip24pt


\maketitle


\begin{sciabstract}
  We present an \textit{ab initio} computational approach for the calculation of resonant Raman intensities,
  including both excitonic and non-adiabatic effects.
  Our diagrammatic approach, which we apply to two prototype, semiconducting layered materials,
  allows a detailed analysis of the impact of phonon-mediated exciton-exciton scattering on the intensities.
  In the case of bulk hexagonal boron nitride, this scattering leads to strong quantum interference between different excitonic resonances,
  strongly redistributing oscillator strength with respect to optical absorption spectra.
  In the case of MoS$_2$, we observe that quantum interference effects are suppressed by the spin-orbit splitting of the excitons.
\end{sciabstract}


\paragraph*{Introduction}

Raman spectroscopy is a versatile tool for the characterization of many materials and, in particular, 2D materials~\cite{tan2018}.
It allows the study of both vibrational and electronic properties.
Among many other things, it serves to probe many-body~\cite{sonntag2018}
and non-adiabatic effects~\cite{pisana2007,yan2007} as well as excitonic ones~\cite{carvalho2015,miranda2017}.
From a theoretical point of view, however, a general and comprehensive approach
that captures both of these effects is missing so far.
Regarding one-phonon Raman scattering, recent theoretical and computational efforts have focused
either on a leading-order perturbation theory approach
~\cite{popov2006b,basko2008,basko2009,venezuela2011,herziger2014,hasdeo2016,reichardt2017b},
which is able to capture non-adiabatic effects but not excitonic ones,
or on a method based on the calculation of the change of the dielectric susceptibility
with static atomic displacements along phonon modes~\cite{knoll1995,ambrosch2002,gillet2013,miranda2017,gillet2017b,wang2018},
which captures excitonic effects but is entirely adiabatic.
As such, an approach that is able to capture both effects is highly desirable not only from a theoretical and conceptional point of view,
but also in practice.
This is especially true for systems with sizable phonon frequencies and strong excitonic effects
or also for materials whose properties are \textit{a priori} unknown.

In a recent work~\cite{reichardt2019}, we laid the groundwork for such a comprehensive theoretical description
of Raman scattering from first principles that allows the inclusive description of both excitonic and non-adiabatic effects
in one fully quantum mechanical theory.
Here we now apply this general theory to the case of resonant, one-phonon Raman scattering.
We present concrete expressions for the Raman intensity in terms of quantities that can be obtained
by \textit{ab initio} calculations on the level of density functional theory (DFT) and many-body perturbation theory
(MBPT), i.e., within the $GW$-approximation for the quasi-particle band structure
and solving the Bethe-Salpeter equation to include excitonic effects.
We then apply it to two materials, bulk hexagonal boron nitride (hBN) and monolayer molybdenum disulfide (MoS$_2$),
which are known to feature strong excitonic effects in both absorption~\cite{arnaud2006,molina2016}
and Raman spectra~\cite{reich2005,carvalho2015} due to their layered, quasi-two-dimensional geometry.
The modular structure of our method allows us to further analyze our numerical results in detail
and explain them in a physically intuitive way.

In the case of hBN our two main findings are (i) the suppression of some of the higher
excitonic resonances in the Raman spectrum, contrary to their brightness in the optical absorption spectrum,
and (ii) strong quantum interference between the first two excitonic resonances mediated by non-adiabatic effects,
which leads to a strong redistribution of scattering weight from the first to the second exciton,
something that cannot be captured in an adiabatic theory.
In the case of monolayer MoS$_2$, we investigate (i) the difference in the coupling of the three main excitonic resonances
to the in-plane and out-out-plane Raman modes and (ii) the effects of non-adiabaticity on the position of the main resonances.

\paragraph{Methods}

Our computational approach to resonant Raman scattering is based on the general theoretical treatment we devised in Ref.~\cite{reichardt2019}.
In this work we are now interested in the special case of one-phonon Raman scattering.
The rate for incident light of frequency $\omega_{\mathrm{L}}$ and polarization direction $\mu$ to scatter
and be detected with frequency $\omega_{\mathrm{D}}$ and polarization $\nu$ after the creation of one phonon is given by:
\begin{equation}
  \dot{P}_{\mathrm{1-ph.}} = \text{const.} \times \omega_{\mathrm{L}} \omega_{\mathrm{D}}
  \sum_{\lambda} \left| \tilde{\mathcal{M}}^{\lambda}_{\mu\nu}(\omega_{\mathrm{L}}-\omega_{\mathrm{D}},\omega_{\mathrm{D}}) \right|^2
                 \mathcal{A}_{\lambda}(\omega_{\mathrm{L}}-\omega_{\mathrm{D}}).
\end{equation}
Here, the sum runs over all phonon modes $\lambda$ with zero momentum, as enforced by momentum conservation
when the light is treated in the dipole approximation, in which the light momentum vanishes.
Eq. (1) expresses the one-phonon part of the Raman spectrum as a sum of phonon spectral functions $\mathcal{A}_{\lambda}$,
which determine the peak shapes, weighted by the scattering matrix element
$\tilde{\mathcal{M}}^{\lambda}_{\mu\nu}(\omega_{\mathrm{L}}-\omega_{\mathrm{D}},\omega_{\mathrm{D}})$,
which determines the peak intensities, or more precisely, their integrated areas.
In this article, we focus on the scattering matrix element alone and treat the phonon spectral function as
being a simple Lorentzian centered on the phonon frequency $\omega_{\lambda}$,
with a width narrow enough to justify the approximation $\omega_{\mathrm{L}} = \omega_{\mathrm{D}} + \omega_{\lambda}$ in the following.
The scattering matrix element contains information on the internal scattering dynamics and
depends on the frequencies of \emph{both} the incident and scattered light.
This is crucial to correctly capture resonant scattering processes between the electronic excitations
inside the material that mediate the scattering process.
In Ref.~\cite{reichardt2019} we also gave a general expression for the scattering matrix element
that mathematically relates it to the correlation of three observables:
an electronic current generated by the incoming light, an electronic current that generates the outgoing light,
and the electronic force on an atom that induces the lattice vibration and creates a phonon.
More precisely, the scattering matrix element is given by the expression
\begin{equation}
  \tilde{\mathcal{M}}^{\lambda}_{\mu\nu}(\omega',\omega) \equiv
  \frac{1}{\hbar^2} \int_{-\infty}^{+\infty} \mathrm{d} t \, \mathrm{e}^{i \omega t}
  \int_{-\infty}^{+\infty} \mathrm{d} t' \, \mathrm{e}^{i \omega' t'}
  (-i) \langle0|\mathcal{T}\left[\hat{F}_{\lambda}(t')\hat{J}_{\nu}(t)\hat{J}^{\dagger}_{\mu}(0)\right]|0\rangle\big|_{\mathrm{connect.}},
\end{equation}
i.e., it is given by the double Fourier transform of a "connected", time-ordered correlation function
of two zero-momentum electronic current densities $\hat{J}_{\mu,\nu}$ and an electronic force on the atoms, $\hat{F}_{\lambda}$
projected on the eigenmode of the phonon to be created.
A precise definition of all the quantities that appear in it is given in Section S1 of the Supplementary Materials (SM).

The advantages of this approach to the description of Raman scattering are twofold.
Firstly, it is a fully dynamical, i.e., it does not neglect any frequency dependence of the intensity as previous approaches do.
This is important when the scattering dynamics involve scattering between electronic states that can be resonantly excited,
not only with the light, but also with the phonon, which cannot be captured with existing, adiabatic approaches.
Naively, these sort of non-adiabatic effects are argued to not be important in wide-gap semiconductors,
as the size of the fundamental energy gap is much larger than typical phonon energies.
However, beyond the fundamental gap, the validity of the adiabatic assumption should not be taken from granted.
Indeed, as we show in the case of the hBN below, the phonon can resonantly couple two \emph{excited} states,
which has an a considerable effect on the Raman intensity.
Secondly, our formulation permits a systematic approximation of the scattering matrix element,
as the correlation function can be controllably approximated with methods from MBPT.
This allows us to derive an expression for the scattering matrix element that includes both intermediate exciton states
and their potentially resonant coupling by a phonon:
\begin{equation}
  \begin{split}
  \tilde{\mathcal{M}}^{\lambda}_{\mu\nu}(\omega_{\lambda},\omega_{\mathrm{L}}-\omega_{\lambda})
  &= \sum_{S,S'} \frac{d^{\mu}_S \big(g^{\lambda}_{S,S'}\big)^* \big(d^{\nu}_{S'}\big)^*}
                      {(\hbar\omega_{\mathrm{L}} - E_S + i\gamma)(\hbar\omega_{\mathrm{L}} - \hbar\omega_{\lambda} - E_{S'} + i\gamma)} \\
  & \quad + \sum_{S,S'} \frac{\big(d^{\mu}_S\big)^* g^{\lambda}_{S,S'} d^{\nu}_{S'}}
                      {(\hbar\omega_{\mathrm{L}} + E_S - i\gamma)(\hbar\omega_{\mathrm{L}} - \hbar\omega_{\lambda} + E_{S'} - i\gamma)}.
  \end{split}
\end{equation}
This expression has the typical form of an expression that appears in third-order quantum mechanical perturbation theory
and describes the excitation of an excitonic state $S$ with energy $E_S$ by the incoming light, its scattering to another excitonic state $S'$
via emission of a phonon, and finally the recombination of exciton $S'$ under emission of light to be detected.
The exciton-light and exciton-phonon matrix elements in the numerators are defined as
\begin{equation}
  d^{\mu}_S \equiv \sum_{\mathbf{k},\substack{a \in \mathcal{C} \\ b \in \mathcal{V}}} d^{\mu,(\mathrm{b})}_{\mathbf{k},a,b} \big(A^S_{\mathbf{k},a,b}\big)^*
\end{equation}
and
\begin{equation}
  g^{\lambda}_{S',S} \equiv \sum_{\mathbf{k}} \bigg[   \sum_{\substack{a,c \in \mathcal{C} \\ b \in \mathcal{V}}}
                                          \big(A^{S'}_{\mathbf{k},a,b}\big)^* g^{\lambda}_{\mathbf{k},a,c} A^S_{\mathbf{k},c,b}
                                        - \sum_{\substack{a \in \mathcal{C} \\ b,c \in \mathcal{V}}}
                                          \big(A^{S'}_{\mathbf{k},a,b}\big)^* g^{\lambda}_{\mathbf{k},c,b} A^S_{\mathbf{k},a,c} \bigg],
\end{equation}
respectively, where $\mathcal{C}$ ($\mathcal{V}$) denotes the set of all conduction (valence) band indices.
$A^S_{\mathbf{k},c,v}$ denotes an excitonic eigenvector, corresponding to an envelope wave function in transition space,
and $d^{\mu,(\mathrm{b})}_{\mathbf{k},a,b}$ and $g^{\lambda}_{\mathbf{k},a,b}$ denote the bare independent-particle dipole moment
and screened electron-phonon coupling matrix elements, respectively.
Finally, $\gamma$ is a broadening factor that we set to a constant value of $75$~meV,
and the sums run over all excitonic states of positive energy.
Details of how to obtain Eq.~(3) from Eq.~(2) are provided in Section~S1 and Figs.~S1 and S2 of the SM.

To make this approach computationally applicable, we implemented Eq.~(3) in a \texttt{Python} code
and obtain the necessary ingredients from first principles calculations.
In general, we obtain the electronic band structure and wave functions on the level of DFT within the local density approximation
in the parametrization due to Perdew and Zunger~\cite{perdew1981}.
The DFT calculations were performed with the \texttt{PWscf} code of the \texttt{QuantumESPRESSO} suite~\cite{giannozzi2009}.
Further material-specific computational details of the DFT calculations are provided in Section S2 of the SM.

The calculation of the excitonic eigenenergies and envelope wave functions was performed with the \texttt{yambo} code~\cite{marini2009}.
The two-particle interaction kernel was computed on the level of the Tamm-Dancoff approximation (TDA) as the
sum of a statically screened, attractive Coulomb part plus a repulsive, bare exchange part.
This approximation is routinely used in calculations of optical absorption spectra~\cite{bechstedt2016,martin2016}
and is sufficient to capture excitonic effects in hBN~\cite{arnaud2006}, MoS$_2$~\cite{molina2016}, and many other materials.
The static inverse dielectric screening function was computed on the level of the random phase approximation
using plane-wave cutoffs of 60~Ry each for the electronic wave functions and the dielectric response function in reciprocal space.
For the calculation of the two-particle interaction kernel, we truncated the attractive screened Coulomb part
at a plane-wave energy cutoff of 10~Ry and the repulsive bare exchange part at a plane-wave cutoff of 20~Ry.
We include a material-specific, finite number of bands in the interaction kernel and
added a rigid energy shift to the independent-particle transition energies
in order to account for the underestimation of the band gap in DFT.
The material specific information is provided in Section S2 of the SM.
Finally, we solve the Bethe-Salpeter equation iteratively for the first 2000 excitonic eigenenergies and envelope wave functions
using the \texttt{SLEPc} library~\cite{hernandez2005}.

For the screened electron-phonon coupling, we approximate the two-particle interaction kernel
on the level of time-dependent density functional theory.
As this approximation is equivalent to obtaining the screened electron-phonon matrix elements
with density functional perturbation theory (see Secs. 4.3 and 4.5 of Ref.~\cite{reichardt2018}),
we calculate the screened matrix elements with the \texttt{PH} code of the \texttt{QuantumESPRESSO} suite~\cite{giannozzi2009}.
For the material specific phonon frequencies, we use the calculated values given in Section S2 of the SM.

\paragraph*{Results and Discussion}

As a first test case, we apply our approach to hexagonal boron nitride in its bulk form with $AA'$ stacking order.
This material features both strong excitonic effects~\cite{arnaud2006} and light atoms,
which leads to relatively large phonon energies of up to $\approx$170~meV.
The latter makes it an ideal candidate to observe the impact
of non-adiabatic effects on the frequency dependence of the resonant Raman intensity.
\begin{figure}[h!tb]
  \centering
  \includegraphics[scale=1]{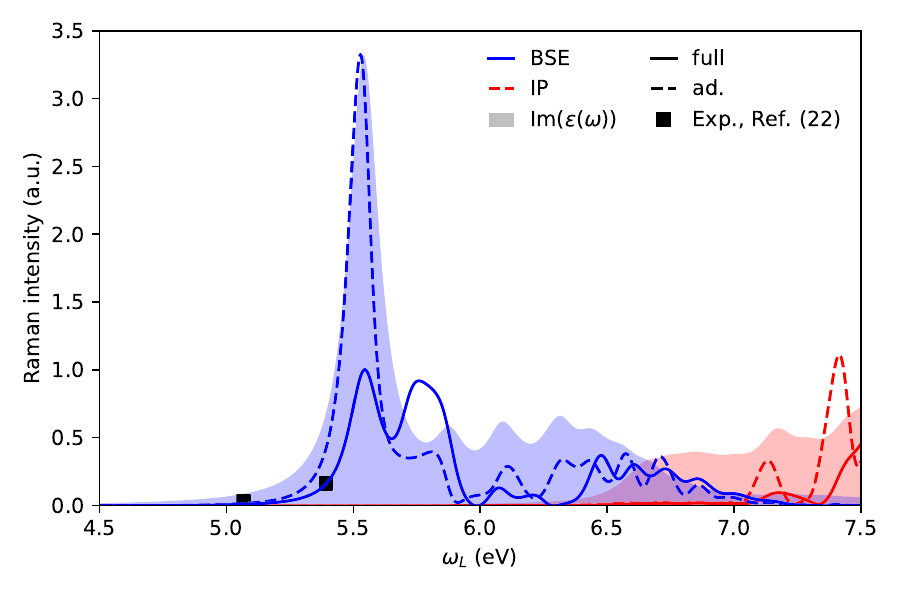}
  \caption{\textbf{Calculated Raman intensities for bulk hexagonal boron nitride as a function of excitation frequency.}
           Blue lines: Raman intensity on the BSE-level calculated with the full non-adiabatic theory (full lines)
                       and in the adiabatic limit (dashed lines).
           Red lines: Same as blue lines, but without excitonic effects, i.e., on the IP-level.
           Shaded areas: Imaginary part of the dielectric function on the BSE- (blue) and IP-levels (red).
           Squares: Experimental data from Ref.~\cite{reich2005}, normalized such that the last point matches the theoretical curve.}
\end{figure}
In Fig.~1, we present the results of our calculation for the Raman-active, degenerate, in-plane optical
($E_{2\mathrm{g}}$) phonon mode, summed over outgoing photon and phonon polarizations and averaged over incoming photon polarizations.
We show the computed Raman intensity as a function of excitation energy
both for the full calculation with excitonic effects (blue lines) and for the independent-particle case (red lines),
where we set the electron-hole interaction to zero while still retaining the TDA.
In addition, we also compute the Raman intensity both in the full, non-adiabatic theory (full lines)
and in the adiabatic limit (dashed lines).
As previously shown~\cite{reichardt2019}, the latter can be obtained from the former
by neglecting the dependence of the scattering matrix element on the phonon frequency.
For comparison, we also show the results of a calculation of the imaginary part of the dielectric function,
which is closely related to the absorption coefficient.
Additionally, we also included two experimental data points, measured in the pre-resonant regime, from Ref.~\cite{reich2005},
which compare decently well to the results of our calculation.

In the IP-picture, the absorption spectrum of bulk hBN (red shade) sets on at the direct gap of around 6.50~eV
located at the $M$-point in the band structure (see top panel of Fig.~2).
When excitonic effects are included (blue shade), the absorption spectrum is dominated by a strong excitonic resonance near 5.58~eV,
followed by a series of excitonic peaks with less oscillator strength (compare also Refs.~\cite{galvani2016,aggoune2018,paleari2018}).
\begin{figure}[h!tb]
  \includegraphics[scale=1]{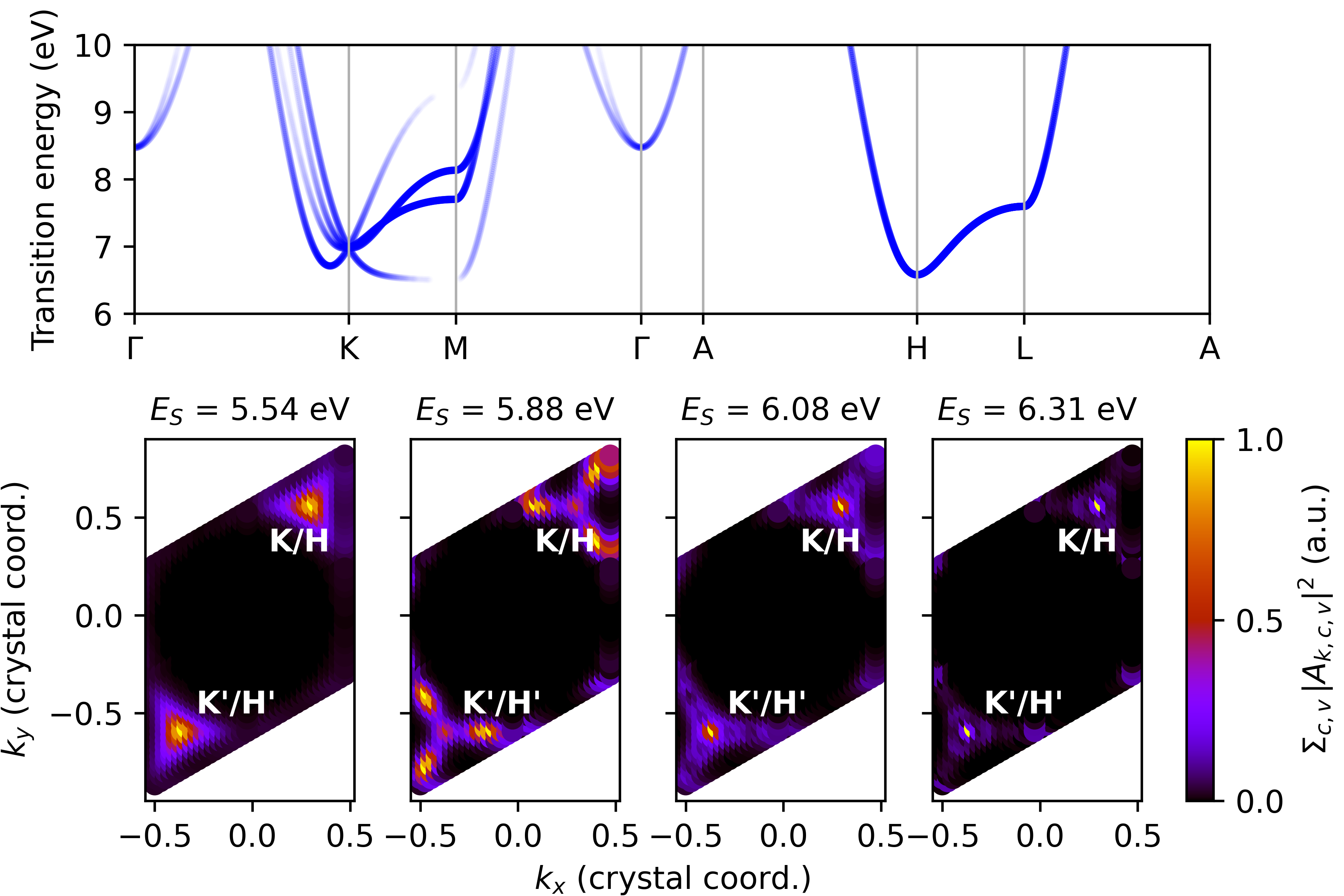}
  \caption{\textbf{hBN: Electronic transition band structure and exciton envelop wave functions in reciprocal space.}
           Top: Electronic transition band structure obtained from the electronic band structure Kohn-Sham DFT within the LDA.
                A rigid shift of 2~eV was added to all conduction band energies.
                The opacity of the lines corresponds to the optical activity of the transition.
           Bottom: Exciton envelope wave functions in reciprocal space for the first four bright excitons.
                   Each panel shows the band-summed square of the envelope wave function $|A^S_{\mathbf{k},c,v}|^2$
                   as a function of the wave vector $\mathbf{k}$ integrated over the $k_z$ coordinate.}
\end{figure}

By contrast, the Raman intensity shows a strikingly different behavior.
On the IP-level, the Raman intensity rises only very slowly after the absorption onset.
This behavior is typical for hexagonal materials and can be traced back to
angular momentum conservation~\cite{reichardt2017,reichardt2017b,miranda2017}.
While the transitions at the direct gap at the $M$-point are relatively suppressed in the Raman process
due to their small dipole moments (see top panel of Fig.~2),
the energetically close transitions around the $H$-point cannot contribute for symmetry reasons.
The latter transitions are located on a circularly symmetric part of the band structure,
which implies full angular momentum conservation.
This in turn suppresses the Raman scattering process for a degenerate phonon,
as the latter carries an effective angular momentum of $\pm \hbar$ and cannot satisfy angular momentum conservation
together with the incoming and outgoing photons.
The same holds true for the transitions around the $K$-point, which also posses full rotation symmetry,
with the exception of a small valley along the $\Gamma$-$K$ direction,
which are the only transitions that can contribute to resonant low-energy Raman scattering on the IP-level.

When excitonic effects are included, we observe a strong suppression of the third and fourth excitonic peak.
This suppression can be understood by analyzing their composition in $\mathbf{k}$-space (see bottom panel of Fig.~2).
The envelope wave functions of these two excitons are strongly localized around the $K$- and $H$-points and are circularly symmetric with only weak trigonally warped contributions.
As a result, they only weakly break the full rotation symmetry around the $K$-and $H$-points in the band structure
and hence the Raman intensity for a degenerate phonon is again strongly suppressed at these frequencies due to angular momentum conservation.
The first two excitons, by contrast, are more delocalized in $\mathbf{k}$-space and their wave functions show strong signs of trigonal warping.
Thus the full rotation symmetry is broken down to the $120^{\circ}$-symmetry of the lattice,
which then allows the Raman process on symmetry grounds.

Even more interesting though is the comparison to the results obtained
within the adiabatic limit (dashed lines in Fig.~1).
In this limit, the relative weight of the first two excitonic resonances follows that of the absorption spectrum.
When non-adiabatic effects are included, however, we observe a strong redistribution of weight from the first to the second exciton.
The reason for this is the non-negligible phonon energy of around 170~meV,
which is close to the energy difference of the first two excitons and leads to strong, resonant inter-exciton scattering.
In order to clearly demonstrate that the redistribution of oscillator strength from the second to the first bright exciton in the non-adiabatic theory indeed arises from inter-exciton scattering, we calculate the Raman intensity with only these contributions.
For this, we define a Raman matrix element that only contains \emph{intra}-exciton scattering processes
via the restricted sum
\begin{equation}
  \begin{split}
  \tilde{\mathcal{M}}^{\lambda}_{\mu\nu}\big|_{\text{intra}}(\omega_{\lambda},\omega_{\mathrm{L}}-\omega_{\lambda})
  =    \sum_{S,S'}\bigg|_{E_S=E_{S'}} \bigg[ & \frac{d^{\mu}_S \big(g^{\lambda}_{S,S'}\big)^* \big(d^{\nu}_{S'}\big)^*}
                       {(\hbar\omega_{\mathrm{L}} - E_S + i\gamma)(\hbar\omega_{\mathrm{L}} - \hbar\omega_{\lambda} - E_{S'} + i\gamma)} \\
    &+ \frac{\big(d^{\mu}_S\big)^* g^{\lambda}_{S,S'} d^{\nu}_{S'}}
                      {(\hbar\omega_{\mathrm{L}} + E_S - i\gamma)(\hbar\omega_{\mathrm{L}} - \hbar\omega_{\lambda} + E_{S'} - i\gamma)} \bigg].
  \end{split}
\end{equation}
We then further define the Raman matrix element with only \emph{inter}-exciton scattering processes as the difference
\begin{equation}
  \tilde{\mathcal{M}}^{\lambda}_{\mu\nu}\big|_{\text{inter}}(\omega_{\lambda},\omega_{\mathrm{L}}-\omega_{\lambda}) \equiv
  \tilde{\mathcal{M}}^{\lambda}_{\mu\nu}(\omega_{\lambda},\omega_{\mathrm{L}}-\omega_{\lambda})
  - \tilde{\mathcal{M}}^{\lambda}_{\mu\nu}\big|_{\text{intra}}(\omega_{\lambda},\omega_{\mathrm{L}}-\omega_{\lambda})
\end{equation}
of the full Raman matrix element of Eq.~(3) and the intra-exciton-only one from Eq.~(6).
Note in this context that many of the exciton states are degenerate.
For example, the first excitonic resonance in the absorption spectrum is due to a bright exciton doublet of $E_{2\mathrm{u}}$-symmetry,
which is slightly higher in energy than the lowest (dark) exciton doublet of $E_{2\mathrm{g}}$-symmetry~\cite{paleari2018}.
The double sum over excitons in Eq.~(6) is therefore understood to run over all exciton states,
taking into account all off-diagonal exciton-phonon matrix elements between degenerate states, but ignoring the coupling of one exciton multiplet to another.

\begin{figure}[h!tb]
  \includegraphics{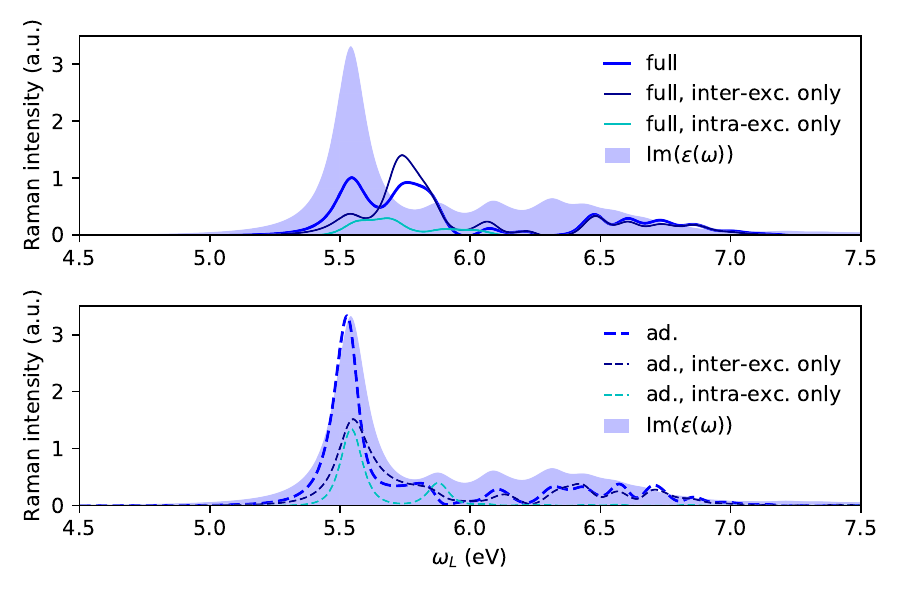}
  \caption{\textbf{hBN: Raman intensity with only inter- or intra-exciton scattering.}
           Raman intensity with all contributions (thick lines) and with only inter- or intra-exciton scattering
           (darker and lighter thin lines, respectively) taken into account.
           Top panel: full, non-adiabatic case.
           Bottom panel: adiabatic limit.
           The shaded area represents the absorption spectrum.
           The Raman intensity is given in the same units in both panels and as in Fig.~1.}
\end{figure}
With these two restricted matrix elements, we can compute the Raman intensity again, this time, however, only with intra- or inter-exciton scattering.
The resulting intensities are shown in Fig.~3.
Firstly, it is important to note that the total intensity is \emph{not} equal to the sum of intensities
with only one of the two scattering channels, as the two contributions are first added to the total matrix element and then squared.
In the case of the adiabatic limit (bottom panel), intra- (lighter color) and inter-exciton scattering (darker color)
contribute to the total Raman intensity with about equal weight and they add up constructively.
By contrast, the full non-adiabatic result (top panel) is strongly dominated by the contribution from \emph{inter-}exciton scattering.
Moreover, while at the first excitonic resonance, inter- and intra-exciton scattering reinforce each other,
at the second excitonic resonance, the two contributions destructively interfere.
Compared to the adiabatic case, we thus observe that the non-adiabatic contribution from the inter-exciton scattering terms changes sign,
which is a typical behavior around a resonance and thus further demonstrates the importance of
considering resonant scattering due to the finite phonon frequency.
This illustrates that non-adiabatic effects can play a significant role even in wide-band gap materials
and emphasizes the importance of considering both non-adiabatic and excitonic effects at the same time.

However, there are also cases in which the resonant, non-adiabatic coupling can be suppressed by other effects.
One example is the case of monolayer molybdenum disulfide which is also known to display strong excitonic effects~\cite{aggoune2018},
yet in which spin-orbit coupling also plays a significant role.
MoS$_2$ features two Raman-active modes, a degenerate in-plane mode of symmetry $E'$
and an out-of-plane one of symmetry $A'_1$.
Furthermore, MoS$_2$ possesses an electronic band structure
that in its optically active region is strongly influenced by spin-orbit coupling~\cite{molina2016}.
In Fig.~4, we show the results of our calculation for both Raman-active modes,
with computational details of the underlying \textit{ab initio} calculations provided in Section S2 of the SM.
The experimental data shown in the figure was taken from Ref.~\cite{carvalho2015}.

\begin{figure}[h!tb]
  \centering
  \includegraphics[scale=1]{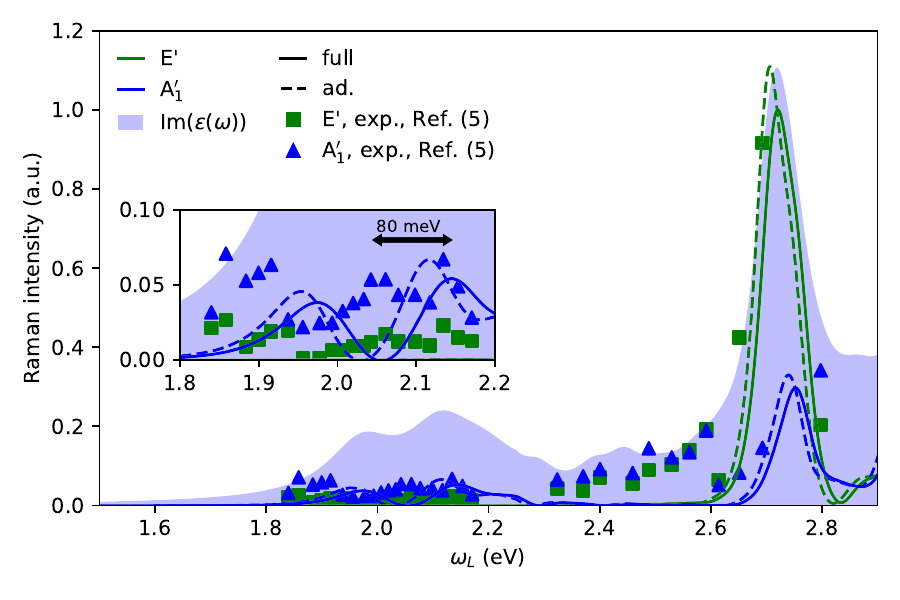}
  \caption{\textbf{Calculated Raman intensities for monolayer molybdenum disulfide as a function of excitation frequency.}
           Green and blue lines: Raman intensity of the $E'$- (green) and $A'_1$-modes (blue) on the BSE-level
                                 calculated with the full non-adiabatic theory (full lines)
                                 and in the adiabatic limit (dashed lines).
           Blue shaded area: Imaginary part of the dielectric function on the BSE-level.
           Green squares and blue triangles: Experimental data from Ref.~\cite{carvalho2015},
           normalized such that the $A'_1$-mode intensity at the second (``B'')-exciton mactches the theoretical curve.
           Inset: Zoom-in into the region around the first two excitons, highlighting an 80 meV shift between theory and experiment.}
\end{figure}
The optical absorption spectrum of monolayer MoS$_2$ (blue shade in Fig.~4) is dominated by three strong excitonic resonances,
at around 1.97~eV, 2.11~eV, and in the region 2.6--2.8~eV.
The first two of these, which each are doubly degenerate, are commonly referred to as the ``A''- and ``B''-exciton~\cite{molina2013,qiu2013b},
while the latter resonance actually consists of several excitons and which are collectively referred to as the ``C''-exciton.
The $\mathbf{k}$-point-resolved envelope wave functions of the ``A''- and ``B''-excitons are strongly localized around the $K$-point
(see bottom panel of Fig.~5) and the difference in energy corresponds to the spin-orbit coupling-induced splitting of the valence band.
By contrast, the dominant ``C''-exciton is strongly delocalized in $\mathbf{k}$-space.
Note that in our calculations, the onset of the ``C''-exciton was obtained at too high an energy.
In order to facilitate a comparison of the relative \emph{heights} of the Raman peaks,
we compressed the excitonic spectrum in excess of 2.25~eV to match the experimental peak positions in absorption spectra.
\begin{figure}[h!tb]
  \includegraphics[scale=1]{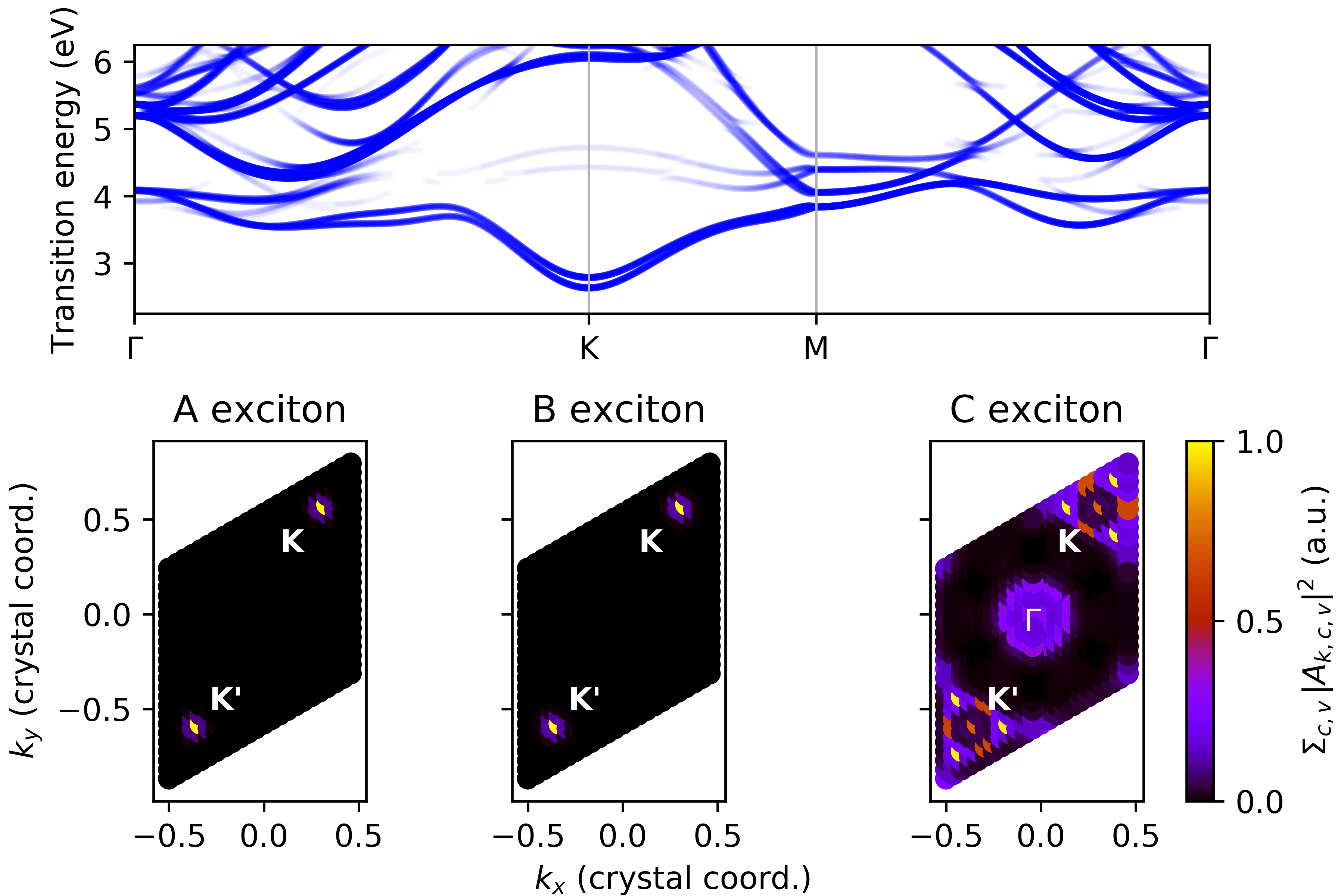}
  \caption{\textbf{MoS$_2$: Electronic transition band structure and exciton envelop wave functions in reciprocal space.}
           Top: Electronic transition band structure obtained from the electronic band structure Kohn-Sham DFT within the LDA.
                A rigid shift of 0.925~eV was added to all conduction band energies.
                The opacity of the lines corresponds to the optical activity of the transition.
           Bottom: Exciton envelope wave functions in reciprocal space for the first three bright excitons
                   (the ``A''-, ``B''-,  and ``C''-excitons).
                   Each panel shows the band-summed square of the envelope wave function $|A^S_{\mathbf{k},c,v}|^2$
                   as a function of the wave vector $\mathbf{k}$.
                   In the case of the ``C''-exciton, we show the sum of the squared wave functions
                   of five energetically close bright excitons.}
\end{figure}

The Raman spectrum of the $A'_1$-mode (blue line) closely follows the trend of the absorption spectrum.
Due to the double-resonant structure of the scattering matrix element, the excitonic peaks become more pronounced and well-separated,
for example, in the case of the ``A'- and ``B''-excitons.
The degenerate $E'$-mode, on the other hand, does not show any resonances at the energies of the ``A'- and ``B''-excitons.
This can again be traced back to their envelope wave functions being strongly localized around the $K$-point,
where the transition band structure exhibits full rotation symmetry (see top panel of Fig.~5),
which in turn suppresses Raman scattering with the $E'$-mode due to angular momentum conservation.
This trend is also seen in the experimental data, in which the $A'_1$-mode features much more prominently in the spectrum.
Note that the peak positions in our calculations appear to be shifted by around 80~meV.
One reason for this could be the missing inclusion of higher-order electron-phonon coupling effects,
such as the zero-point vibrational renormalization of the electronic band structure,
which is known to reduce the fundamental band gap by around 75~meV~\cite{molina2016}.
In the region around the ``C''-exciton, however, the excitons couple much more strongly to the $E'$-mode than to the $A'_1$-mode.
This behavior can again be understood from an analysis in terms of inter- and intra-exciton scattering (see Fig.~6).

\begin{figure}[h!tb]
  \includegraphics{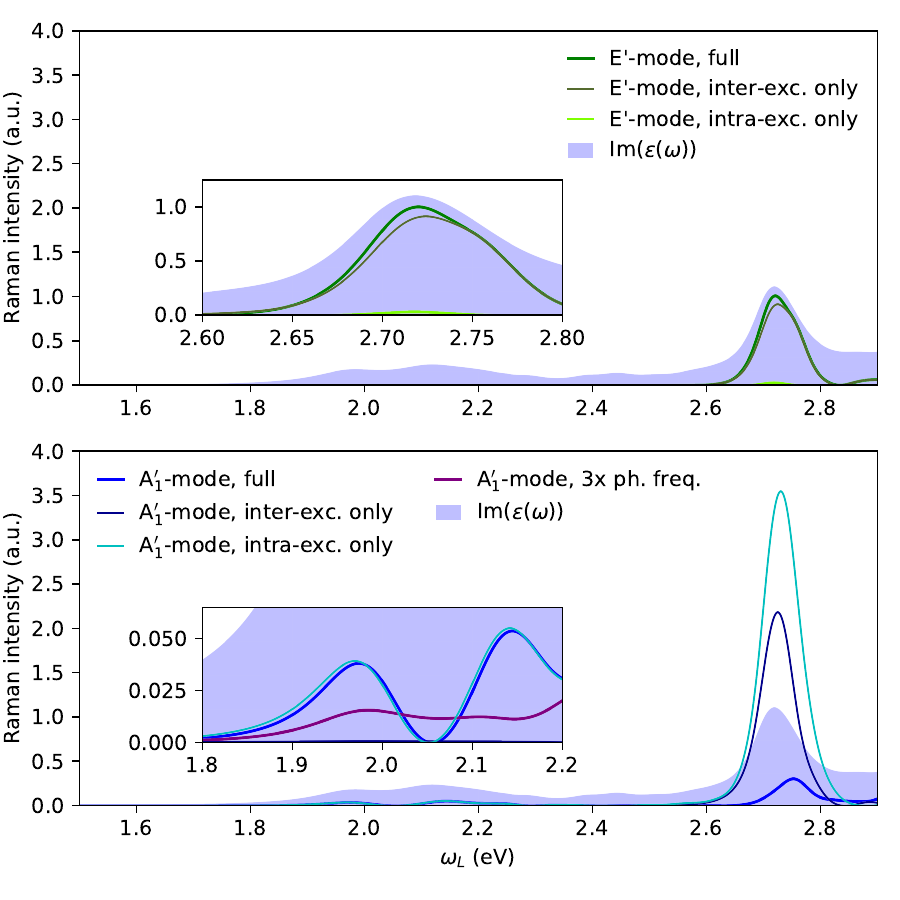}
  \caption{\textbf{MoS$_2$: Raman intensity with only inter- or intra-exciton scattering.}
           Raman intensity with all contributions (thick lines) and with only inter- or intra-exciton scattering
           (darker and lighter thin lines, respectively) taken into account.
           Top panel: $E'$-mode.
           Bottom panel: $A'_1$-mode.
           The shaded area represents the absorption spectrum.
           The Raman intensity is given in the same units in both panels and as in Fig.~4.
           Insets: Zoom-ins into the areas around the three dominant excitonic resonances.
           In the bottom inset, the purple line depicts the Raman intensity of the $A'_1$-mode with an artificially tripled phonon frequency.}
\end{figure}
In Fig.~6, we show the Raman intensity with only one of the two contributions
for both the $E'$-mode (top panel) and the $A'_1$-mode (bottom mode).
We only show the results for the non-adiabatic theory, as the adiabatic limit yields almost the same results.

Firstly, we note that in case of the $E'$-mode, the Raman intensity is strongly suppressed around the ``A''- and ``B''-excitons,
irrespective of whether only intra- or inter-exciton scattering is considered alone.
This further underlines that the suppression is indeed the result of a physical symmetry,
i.e., in this case approximate full angular momentum conservation.
Secondly, the Raman intensity for the $E'$-mode in the energy regime close to the ``C''-exciton is dominated inter-exciton scattering
and intra-exciton scattering plays almost no role.
In the case of the $A'_1$-mode, the picture is entirely different, however.
The low-energy regime around the ``A''- and ``B''-excitons is governed by intra-exciton scattering due to the isolated nature of the two excitons.
At higher excitation energies, inter-exciton scattering gains in weight and becomes almost as important as intra-exciton processes.
Most importantly though is the fact that the two scattering mechanism contribute to the total Raman matrix element with opposite signs.
As such, the lower intensity of the $A'_1$-mode compared to that of the $E'$-mode is the effect of destructive quantum interference.

This results in the $E'$-mode featuring a much larger intensity than the $A'_1$-mode in the exciton continuum.
By contrast, the isolated nature of the ``A''- and ``B''-excitons means that inter-exciton scattering plays no role
and only scattering within either the ``A''- or ``B''-exciton doublet is important.
This analysis also corrects the controversial statement of the recent Ref.~\cite{wang2018}
that only inter-exciton scattering should be considered in the resonant regime, i.e.,
the ``A''- and ``B''-excitons should be silent.
This is not only in contrast to the analysis presented here, but is also in stark contrast to experiments
of other monolayer transition metal dichalcogenites~\cite{miranda2017}.

Finally, we comment on the effects of non-adiabaticity in the case of MoS$_2$.
Here, the phonon energies of the Raman active modes are on the order of 50~meV~\cite{molina2011}.
This is not enough to overcome the spin-orbit coupling-induced splitting of the ``A''- and ``B''-excitons of around 110~meV.
The contributions from the ``A''- and ``B''-excitonic resonances to the scattering matrix element
are thus prevented from interfering with one another.

To verify this hypothesis, we artificially triple the phonon frequency
to overcome the gap between the ``A''-to-``B'' excitons and computed the Raman intensity again
(see purple line in inset in bottom panel of Fig.~6).
Now the intensity decreases significantly as a result of destructive interference between the two excitonic resonances,
which are now resonantly coupled.
We can thus conclude that in the real case, the spin-orbit splitting suppresses the resonant coupling of the two excitons
with a phonon, leading to a bigger Raman intensity.
As such, the inclusion of non-adiabatic effects only results in a slight blueshift of all resonances
compared to the adiabatic case (dashed lines in Fig.~4).

In conclusion, we have presented a method for the computation of resonant one-phonon
Raman intensities from first principles that takes into account both excitonic and non-adiabatic effects.
We have applied our method to both bulk hexagonal boron nitride and single-layer molybdenum disulfide.
In the case of the former, we have explained the absence of particular higher excitonic resonances
from the Raman intensity spectrum with angular momentum conservation.
More importantly, we have proven the significance of non-adiabaticity for the Raman process,
as it leads to strong quantum interference between the first two excitonic resonances.
For MoS$_2$, we have shown that spin-orbit splitting protects the lowest two excitonic resonances from non-adiabatic effects
and that the in-plane $E'$-mode is silent in this regime due to angular momentum conservation.
Furthermore, we have explained the different behavior of the two Raman active modes at higher excitation energies
by destructive quantum interference of intra- and inter-exciton scattering processes involving the out-of-plane $A'_1$-mode.


%

\paragraph*{Acknowledgments}

The authors acknowledge the use of the \texttt{QuantumESPRESSO} suite~\cite{giannozzi2009}
for the DFT calculations within the local density approximation~\cite{perdew1981},
the \texttt{yambo} code~\cite{marini2009} and the \texttt{SLEPc}~\cite{hernandez2005} library for the MBPT calculations,
and usage of the HPC facilities of the University of Luxembourg~\cite{varrette2014}.
All data needed to evaluate the conclusions in the paper are present in the paper and/or the Supplementary Materials.
Additional data available from authors upon request.
The authors would like to thank A.~Marini for inspiring discussions.
S.R. and L.W. acknowledge financial support by the National Research Fund (FNR) Luxembourg
(projects RAMGRASEA and INTER/ANR/13/20/NANOTMD).
The authors declare that they have no competing interests.
Author contributions: S.R. and L.W. devised the project idea.
S.R. developed the theory, implemented it, performed the calculations, and prepared the main part of the manuscript.
Both authors discussed the results, the ideas for their analysis, and edited the manuscript.

\paragraph*{Supplementary materials}
The supplementary materials include the following items:\\
Section S1 -- Theoretical details\\
Figure S1 -- Diagrammatic analysis of the scattering matrix element \\
Figure S2 -- Ladder-like approximation to the scattering matrix element\\
Section S2 -- Computational details for the hBN and MoS$_2$ calculations



\section*{Supplementary material}

# Non-adiabatic exciton-phonon coupling in Raman spectroscopy of layered materials – Supplementary Material

Sven Reichardt[1] and Ludger Wirtz[1]

[1]*Department of Physics and Materials Science, University of Luxembourg, 1511 Luxembourg, Luxembourg*

The Supplementary Material is structured as follows: In Section S1, we provide the theoretical details underlying our approach. Section S2 contains material-specific details of the first-principles calculations.

## S1. THEORETICAL DETAILS

Here we provide more details on how to obtain Eq. (3) from Eq. (2) of the main text.
We start from Eq. (2) of the main text,

$$\tilde{\mathcal{M}}^\lambda_{\mu\nu}(\omega',\omega) \equiv \frac{1}{\hbar^2} \int_{-\infty}^{+\infty} \mathrm{d}t\, e^{i\omega t} \int_{-\infty}^{+\infty} \mathrm{d}t'\, e^{i\omega' t'} (-i)\langle 0|\mathcal{T}\left[\hat{F}_\lambda(t')\hat{J}_\nu(t)\hat{J}^\dagger_\mu(0)\right]|0\rangle\big|_{\mathrm{connect.}}, \quad (S1)$$

and first provide a definition of all the appearing quantities before proceeding with the derivation of Eq. (3) of the main text. The correlation function on the right-hand side involves: (i) the spatially Fourier-transformed electronic current density operator $\hat{J}$ in the dipole approximation, projected onto the light polarization vector $\boldsymbol{\epsilon}_\mu$:

$$\hat{J}_\mu \equiv (-e/m)\boldsymbol{\epsilon}^*_\mu \cdot \int \mathrm{d}^3 r\, \hat{\psi}^\dagger(\mathbf{r})(-i\hbar\boldsymbol{\nabla})\hat{\psi}(\mathbf{r}) \quad (S2)$$

and (ii) the electronic force operator $\hat{F}_\lambda$, projected onto the eigenvector of a zero-momentum phonon mode $\lambda$:

$$\hat{F}_\lambda \equiv \sum_I \mathbf{v}^\lambda_I \cdot \int \mathrm{d}^3 r\, \hat{\psi}^\dagger(\mathbf{r})\hat{\psi}(\mathbf{r}) \partial V_{\mathrm{lat}}(\mathbf{r})/\partial(\mathbf{R}_I)\big|_{\{\mathbf{R}^{(0)}_I\}}. \quad (S3)$$

In the latter, the sum over $I$ runs over all atoms and $\mathbf{v}^\lambda_I$ denotes the displacement vector (with dimensions of length) of atom $I$ according to the zero-momentum phonon mode $\lambda$. Further, $V_{\mathrm{lat}}(\mathbf{r})$ denotes a fixed lattice potential of nuclei with equilibrium configuration $\{\mathbf{R}^{(0)}_I\}$ and $\hat{\psi}(\mathbf{r})$ is the electron field operator, with $(-e)$ and $m$ being the charge and mass of an electron, respectively. Finally, all time-dependent operators are understood to be in the Heisenberg picture with respect to the full Hamiltonian of the matter system, whose ground state is denoted by $|0\rangle$.

The challenge in making the scattering matrix element of Eq. (S1) accessible to a numerical computation lies in the complexity of the three-particle correlation function appearing as part of the force-current-current correlation function. In order to find an approximation for it that still captures excitonic and non-adiabatic effects, we treat both the electron-electron and the electron-phonon interaction in perturbation theory and carry out a diagrammatic analysis. This allows us to systematically truncate the perturbation series in a way that captures the relevant physics, but can still be evaluated in a way suitable for a numerical calculation.

To start with, we expand the bare (b) electronic current density and force operators in the basis of Kohn-Sham states:

$$\hat{J}_\mu \equiv \sum_{\mathbf{k},a,a'} \left(d^{\mu,(\mathrm{b})}_{\mathbf{k},a',a}\right)^* \hat{c}^\dagger_{\mathbf{k},a}\hat{c}_{\mathbf{k},a'} \quad (S4)$$

$$\hat{F}_\lambda \equiv \sum_{\mathbf{k},b,b'} \left(g^{\lambda,(\mathrm{b})}_{\mathbf{k},b',b}\right)^* \hat{c}^\dagger_{\mathbf{k},b}\hat{c}_{\mathbf{k},b'}. \quad (S5)$$

The expansion coefficients are defined such that they denote the matrix elements for *photon/phonon emission*, i.e., *current density/force annihilation*. Note that at zero momentum transfer and for real polarization vectors/phonon eigenvectors, the matrix elements obey $m^{(\mathrm{b})}_{\mathbf{k},a',a} = \left(m^{(\mathrm{b})}_{\mathbf{k},a,a'}\right)^*$.

In terms of these matrix elements, the time-ordered force-current-current correlation function reads

$$\langle 0|\mathcal{T}\left[\hat{F}_\lambda(t')\hat{J}_\nu(t)\hat{J}^\dagger_\mu(0)\right]|0\rangle\big|_{\mathrm{connect.}} = \sum_{\mathbf{k},\mathbf{k}',\mathbf{k}''} \sum_{\substack{a,b,c \\ a',b',c'}} \left(g^{\lambda,(\mathrm{b})}_{\mathbf{k},a',a}\right)^* \left(d^{\nu,(\mathrm{b})}_{\mathbf{k}',b',b}\right)^* d^{\mu,(\mathrm{b})}_{\mathbf{k}'',c,c'}$$
$$\times \langle 0|\mathcal{T}\left[\hat{c}^\dagger_{\mathbf{k},a}(t')\hat{c}_{\mathbf{k},a'}(t')\hat{c}^\dagger_{\mathbf{k}',b}(t)\hat{c}_{\mathbf{k}',b'}(t)\hat{c}^\dagger_{\mathbf{k}'',c}(0)\hat{c}_{\mathbf{k}'',c'}(0)\right]|0\rangle\big|_{\mathrm{connect.}}. \quad (S6)$$

2

The last factor is the fully connected part of the three-particle Green's function. We approximate it in a controlled way with the help of Feynman diagrams, as shown in Fig. S1. The first term shown corresponds to the independent-particle

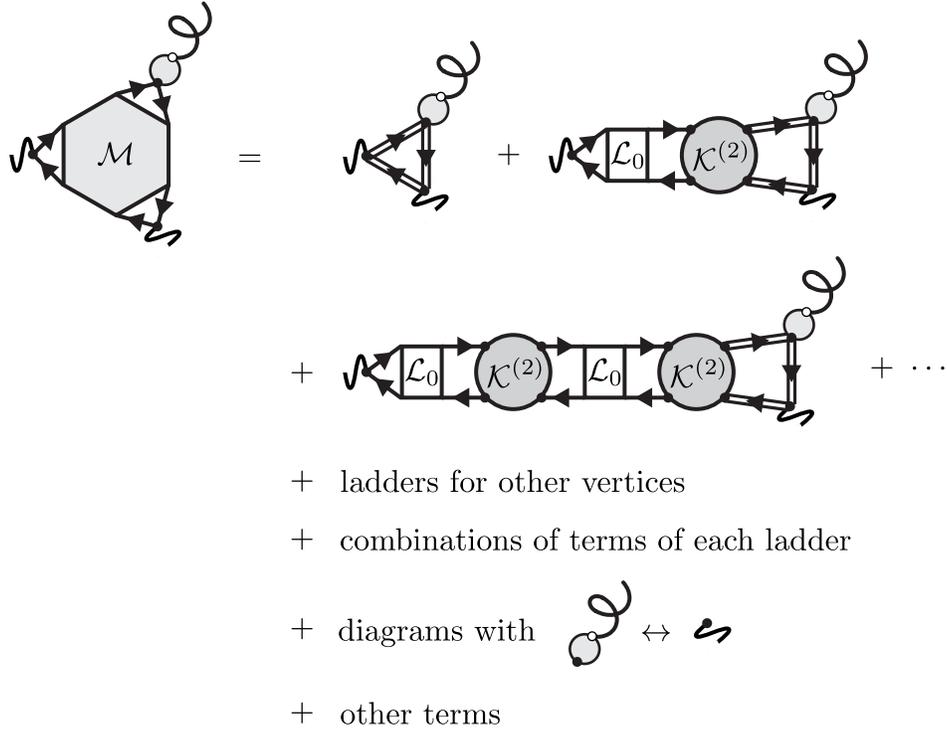

+ ladders for other vertices

+ combinations of terms of each ladder

+ diagrams with ↔

+ other terms

FIG. S1. **Diagrammatic analysis of the scattering matrix element.** Dark dots (light circles) denote the bare electron-photon (-phonon) vertex. Double lines denote the one-electron Green's function. $\mathcal{L}$ ($\mathcal{L}_0$) and $\mathcal{K}^{(2)}$ denote the interacting (independent-particle) two-particle correlation function and two-particle interaction kernel, respectively.

approximation to the three-particle correlation function. In the second term, we include the first-order correction to the electron-incoming photon vertex due to two-particle inter-electron interactions. We denote the kernel describing the latter by $\mathcal{K}^{(2)}$, while $\mathcal{L}_0$ represents the two-particle correlation function on the independent-particle level. The second line shows the second-order correction to the same vertex while the dots refer to higher-order corrections of the same form. We consider the same summation of terms for all three vertices simultaneously. In order to consider all topologically distinct diagrams, we also need to consider terms in which the fermion flow is reversed, i.e., in which the vertices for the outgoing photon and phonon are interchanged.

We do now adopt what we refer to as the *ladder-like approximation* in the following. It consists of neglecting all terms that do not follow the ladder-like structure depicted in the first two lines of Fig. S1. As shown in Fig. S2, the summation of terms at each vertex can be carried out separately. The double Fourier transform of the force-current-current correlation function, which up to a factor of $i$ corresponds to the scattering matrix element, then takes on the form

$$i\tilde{\mathcal{M}}^\lambda_{\mu\nu}(\omega',\omega) = \sum_{\mathbf{k}} \sum_{\substack{a,b,c \\ a',b',c'}} \overline{g}^\lambda_{\mathbf{k},a',a}(\omega')\overline{d}^\nu_{\mathbf{k},b',b}(\omega)d^\mu_{\mathbf{k},c,c'}(\omega+\omega') \\ \times \frac{1}{\hbar^2} \int_{-\infty}^{+\infty} dt\, e^{i\omega t} \int_{-\infty}^{+\infty} dt'\, e^{i\omega' t'} \langle 0|\mathcal{T}\left[\hat{c}^\dagger_{\mathbf{k},a}(t')\hat{c}_{\mathbf{k},a'}(t')\hat{c}^\dagger_{\mathbf{k},b}(t)\hat{c}_{\mathbf{k},b'}(t)\hat{c}^\dagger_{\mathbf{k},c}(0)\hat{c}_{\mathbf{k},c'}(0)\right]|0\rangle\big|_{\text{IP,connect.}},$$
(S7)

where the three-particle correlation function now only appears in the form of its independent-particle version. The first three factors denote the Fourier transforms of the *exact* vertices, with a bar denoting the vertices for photon/phonon emission. To obtain concrete expressions for the latter, we approximate the two-particle interaction kernel as described in the following.

In the case of the electron-light vertex, we approximate it as the sum of a statically screened, attractive electron-hole Coulomb interaction and a bare, repulsive exchange part. We can then give an explicit expression for the

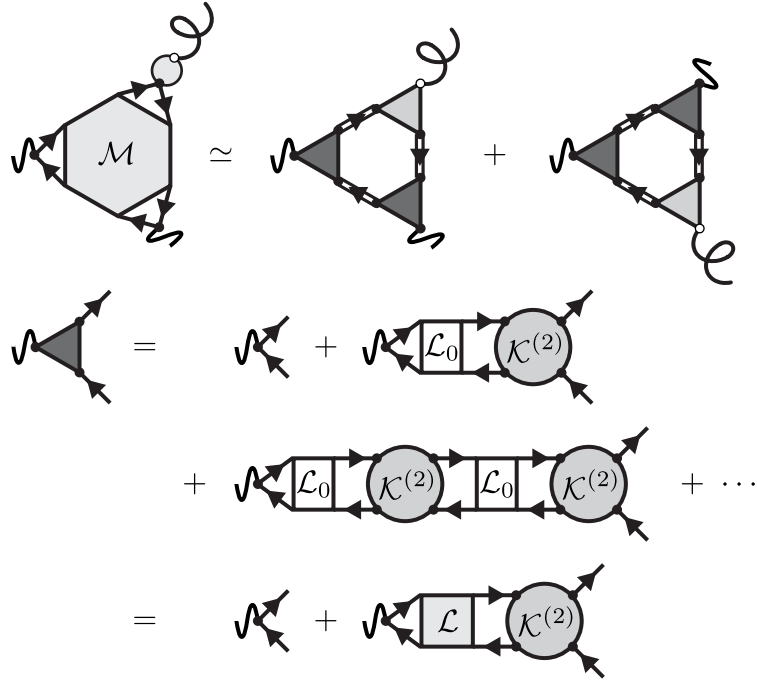

FIG. S2. **Ladder-like approximation to the scattering matrix element.** Top: Diagrammatic approximation for the scattering matrix element (left) as a sum of two terms with effectively opposite orientation of the fermion loop. Dark dots (light circles) denote the bare electron-photon (-phonon) vertex. Double lines denote the one-electron Green's function. Dark (light) shaded triangles the exact electron-photon (-phonon) vertex. Bottom: Exact ladder of diagrams for the electron-light coupling. $\mathcal{L}$ ($\mathcal{L}_0$) and $\mathcal{K}^{(2)}$ denote the interacting (independent-particle) two-particle correlation function and two-particle interaction kernel, respectively. Note that the same summation is understood to be carried out at each vertex.

Fourier-transformed vertex function by making use of the Bethe-Salpeter equation

$$L_{\mathbf{k},a',a}(\omega) = L_{0;\mathbf{k},a',a}(\omega)\delta_{\mathbf{k},\mathbf{k}'}\delta_{a',b'}\delta_{a,b} + L_{0;\mathbf{k},a',a}(\omega)\sum_{\substack{\mathbf{k}''\\ \mathbf{k}',b',b}}\sum_{\substack{c',c\\ \mathbf{k}'',c',c}} K^{(2)}_{\substack{\mathbf{k},a',a\\ \mathbf{k}'',c',c}} L_{\substack{\mathbf{k}'',c',c\\ \mathbf{k}',b',b}}(\omega) \tag{S8}$$

where we now use non-calligraphic script to denote the approximated quantities. Note that we also use the quasi-particle approximation for $L_0$, so that it only depends on two band indices. As shown in Fig. S2, we can identify the right-hand side up to a factor of $L_0$ as the factor that modulates the bare vertex, and we thus have

$$d^\mu_{\mathbf{k},a',a}(\omega) = d^{\mu,(b)}_{\mathbf{k},a',a} + \sum_{\substack{\mathbf{k}',\mathbf{k}''\\ b',c'\\ b,c}}\sum K^{(2)}_{\substack{\mathbf{k},a',a\\ \mathbf{k}',b',b}} L_{\substack{\mathbf{k}',b',b\\ \mathbf{k}'',c',c}}(\omega) d^{\mu,(b)}_{\mathbf{k}'',c',c} = L^{-1}_{0;\mathbf{k},a',a}(\omega)\sum_{\mathbf{k}'}\sum_{\substack{b',b\\ \mathbf{k}',b',b}} L_{\substack{\mathbf{k},a',a\\ \mathbf{k}',b',b}}(\omega) d^{\mu,(b)}_{\mathbf{k}',b',b} \tag{S9}$$

for photon *absorption* and

$$\overline{d}^\mu_{\mathbf{k},a',a}(\omega) = \left(d^{\mu,(b)}_{\mathbf{k},a',a}\right)^* + \sum_{\substack{\mathbf{k}',\mathbf{k}''\\ b',c'\\ b,c}}\sum \left(d^{\mu,(b)}_{\mathbf{k}',b',b}\right)^* L_{\substack{\mathbf{k}',b',b\\ \mathbf{k}'',c',c}}(\omega) K^{(2)}_{\substack{\mathbf{k}'',c',c\\ \mathbf{k},a',a}} = \sum_{\mathbf{k}'}\sum_{b',b}\left(d^{\mu,(b)}_{\mathbf{k}',b',b}\right)^* L_{\substack{\mathbf{k}',b',b\\ \mathbf{k},a',a}}(\omega) L^{-1}_{0;\mathbf{k},a',a}(\omega). \tag{S10}$$

for photon *emission*. Lastly, as mentioned in the main text, we also use the *Tamm-Dancoff approximation*, which is justified for the semiconducting systems discussed in the main text. In this approximation, the independent-particle two-particle correlation function reads

$$L_{0;\mathbf{k},a',a}(\omega) = \frac{\bar{f}_{\mathbf{k},a'} f_{\mathbf{k},a}}{\hbar\omega - (\varepsilon_{\mathbf{k},a'} - \varepsilon_{\mathbf{k},a}) + i\eta} + \frac{f_{\mathbf{k},a'} \bar{f}_{\mathbf{k},a}}{-\hbar\omega - (\varepsilon_{\mathbf{k},a} - \varepsilon_{\mathbf{k},a'}) + i\eta}, \tag{S11}$$

where $f_{\mathbf{k},a}$ denotes the occupancy of the state $|\mathbf{k},a\rangle$ in the ground state, $\bar{f}_{\mathbf{k},a} \equiv 1 - f_{\mathbf{k},a}$, and $\eta$ is a positive infinitesimal. The interacting two-particle correlation function can be expressed in terms of the excitonic wave functions $A^S_{\mathbf{k},a',a}$ and energies $E_S$ as

$$L_{\mathbf{k},a',a}(\omega) = \sum_{\substack{S \\ \mathbf{k}',b',b}} \left[ \bar{f}_{\mathbf{k},a'} f_{\mathbf{k},a} \bar{f}_{\mathbf{k}',b'} f_{\mathbf{k}',b} \frac{A^S_{\mathbf{k},a',a} \left(A^S_{\mathbf{k}',b',b}\right)^*}{\hbar\omega - E_S + i\eta} + f_{\mathbf{k},a'} \bar{f}_{\mathbf{k},a} f_{\mathbf{k}',b'} \bar{f}_{\mathbf{k}',b} \frac{\left(A^S_{\mathbf{k},a,a'}\right)^* A^S_{\mathbf{k}',b,b'}}{-\hbar\omega - E_S + i\eta} \right], \quad \text{(S12)}$$

where the sum runs over all *positive-energy* excitons. Note that within the Tamm-Dancoff approximation, an excitonic envelope wave function $A^S_{\mathbf{k},a',a}$ for positive-energy excitons has non-vanishing components only for $a' \in \mathcal{C}$ and $a \in \mathcal{V}$, where $\mathcal{C}$ and $\mathcal{V}$ denote the sets of conduction and valence band indices, respectively. Combining the last three equations, we arrive at the final expression we use to approximate the exact electron-photon vertices:

$$d^\mu_{\mathbf{k},a',a}(\omega) = \sum_S \left[ \bar{f}_{\mathbf{k},a'} f_{\mathbf{k},a} A^S_{\mathbf{k},a',a} \frac{\hbar\omega - (\varepsilon_{\mathbf{k},a'} - \varepsilon_{\mathbf{k},a}) + i\eta}{\hbar\omega - E_S + i\eta} d^\mu_S \right.$$
$$\left. + f_{\mathbf{k},a'} \bar{f}_{\mathbf{k},a} \left(A^S_{\mathbf{k},a,a'}\right)^* \frac{-\hbar\omega - (\varepsilon_{\mathbf{k},a} - \varepsilon_{\mathbf{k},a'}) + i\eta}{-\hbar\omega - E_S + i\eta} (d^\mu_S)^* \right] \quad \text{(S13)}$$

and

$$\bar{d}^\mu_{\mathbf{k},a',a}(\omega) = \sum_S \left[ \bar{f}_{\mathbf{k},a'} f_{\mathbf{k},a} (d^\mu_S)^* \frac{\hbar\omega - (\varepsilon_{\mathbf{k},a'} - \varepsilon_{\mathbf{k},a}) + i\eta}{\hbar\omega - E_S + i\eta} \left(A^S_{\mathbf{k},a',a}\right)^* \right.$$
$$\left. + f_{\mathbf{k},a'} \bar{f}_{\mathbf{k},a} d^\mu_S \frac{-\hbar\omega - (\varepsilon_{\mathbf{k},a} - \varepsilon_{\mathbf{k},a'}) + i\eta}{-\hbar\omega - E_S + i\eta} A^S_{\mathbf{k},a,a'} \right], \quad \text{(S14)}$$

with $d^\mu_S$ as defined in Eq. (4) of the main text.

For the exact electron-phonon vertex, we follow the same line of argument, but approximate the two-particle interaction kernel in that case with the kernel corresponding to time-dependent density functional theory. As shown in Secs. 4.3 and 4.5 of Ref. (29), this approximation is exactly equivalent to using the statically screened electron-phonon coupling on the level of density functional perturbation theory. In addition, we also neglect the frequency dependence of the vertex function, i.e., we use

$$\bar{g}^\lambda_{\mathbf{k},a',a}(\omega) \to \left(g^\lambda_{\mathbf{k},a',a}\right)^* \big|_{\text{DFPT}}, \quad \text{(S15)}$$

which is justified for systems with a band gap much larger than the phonon frequency. Note, however, that we *do retain the phonon frequency-dependence of the three-particle independent-particle correlation function* as there the frequency of the photon needs to be considered as well, which reduces the effective band gap seen by the phonon in the three-particle correlation function.

Finally, we also give the explicit form of the Fourier transformed independent-particle three-particle correlation function. Treating it on the level of the quasi-particle approximation, one finds

$$\frac{1}{\hbar^2} \int_{-\infty}^{+\infty} dt\, e^{i\omega t} \int_{-\infty}^{+\infty} dt'\, e^{i\omega' t'} \langle 0|\mathcal{T}\left[\hat{c}^\dagger_{\mathbf{k},a}(t')\hat{c}_{\mathbf{k},a'}(t')\hat{c}^\dagger_{\mathbf{k},b}(t)\hat{c}_{\mathbf{k},b'}(t)\hat{c}^\dagger_{\mathbf{k},c}(0)\hat{c}_{\mathbf{k},c'}(0)\right]|0\rangle\big|_{\text{IP,connect.}}$$
$$= \frac{1}{\hbar^2} \int_{-\infty}^{+\infty} dt\, e^{i\omega t} \int_{-\infty}^{+\infty} dt'\, e^{i\omega' t'} (-1)$$
$$\times \left\{ \delta_{a',b}\delta_{b',c}\delta_{c',a} \langle 0|\mathcal{T}\left[\hat{c}_{\mathbf{k},a}(t')\hat{c}^\dagger_{\mathbf{k},a}(t)\right]|0\rangle \langle 0|\mathcal{T}\left[\hat{c}_{\mathbf{k},b}(t)\hat{c}^\dagger_{\mathbf{k},b}(0)\right]|0\rangle \langle 0|\mathcal{T}\left[\hat{c}_{\mathbf{k},c}(0)\hat{c}^\dagger_{\mathbf{k},c}(t')\right]|0\rangle \right.$$
$$\left. + \delta_{a',c}\delta_{c',b}\delta_{b',a} \langle 0|\mathcal{T}\left[\hat{c}_{\mathbf{k},a}(t')\hat{c}^\dagger_{\mathbf{k},a}(0)\right]|0\rangle \langle 0|\mathcal{T}\left[\hat{c}_{\mathbf{k},b}(t)\hat{c}^\dagger_{\mathbf{k},b}(t')\right]|0\rangle \langle 0|\mathcal{T}\left[\hat{c}_{\mathbf{k},c}(0)\hat{c}^\dagger_{\mathbf{k},c}(t)\right]|0\rangle \right\} \quad \text{(S16)}$$
$$\equiv i\,\delta_{a',b}\delta_{b',c}\delta_{c',a} R_{0;\mathbf{k},c,a,b}(-\omega',-\omega) + i\,\delta_{a',c}\delta_{c',b}\delta_{b',a} R_{0;\mathbf{k},a,b,c}(\omega',\omega).$$

In the last step, we identified the Fourier transform of the independent-particle three-particle correlation function on

the quasi-particle level, which is explicitly given by

$$
\begin{aligned}
&(-i)R_{0;\mathbf{k},a,b,c}(\omega_1,\omega_2)\\
&=\frac{1}{\hbar^2}\int_{-\infty}^{+\infty}\mathrm{d}t\,e^{i\omega t}\int_{-\infty}^{+\infty}\mathrm{d}t'\,e^{i\omega' t'}\langle 0|\mathcal{T}\left[\hat{c}_{\mathbf{k},a}(t')\hat{c}_{\mathbf{k},a}^{\dagger}(0)\right]|0\rangle\langle 0|\mathcal{T}\left[\hat{c}_{\mathbf{k},b}(t)\hat{c}_{\mathbf{k},b}^{\dagger}(t')\right]|0\rangle\langle 0|\mathcal{T}\left[\hat{c}_{\mathbf{k},c}(0)\hat{c}_{\mathbf{k},c}^{\dagger}(t)\right]|0\rangle\\
&=+\frac{f_{\mathbf{k},a}\bar{f}_{\mathbf{k},b}\bar{f}_{\mathbf{k},c}}{[-\hbar\omega_1-\hbar\omega_2-(\varepsilon_{\mathbf{k},c}-\varepsilon_{\mathbf{k},a})+i\eta][-\hbar\omega_1-(\varepsilon_{\mathbf{k},b}-\varepsilon_{\mathbf{k},a})+i\eta]}\\
&\quad-\frac{\bar{f}_{\mathbf{k},a}f_{\mathbf{k},b}f_{\mathbf{k},c}}{[\hbar\omega_1+\hbar\omega_2-(\varepsilon_{\mathbf{k},a}-\varepsilon_{\mathbf{k},c})+i\eta][\hbar\omega_1-(\varepsilon_{\mathbf{k},a}-\varepsilon_{\mathbf{k},b})+i\eta]}\\
&\quad+\frac{\bar{f}_{\mathbf{k},a}\bar{f}_{\mathbf{k},b}f_{\mathbf{k},c}}{[\hbar\omega_1+\hbar\omega_2-(\varepsilon_{\mathbf{k},a}-\varepsilon_{\mathbf{k},c})+i\eta][\hbar\omega_2-(\varepsilon_{\mathbf{k},b}-\varepsilon_{\mathbf{k},c})+i\eta]}\\
&\quad-\frac{f_{\mathbf{k},a}f_{\mathbf{k},b}\bar{f}_{\mathbf{k},c}}{[-\hbar\omega_1-\hbar\omega_2-(\varepsilon_{\mathbf{k},c}-\varepsilon_{\mathbf{k},a})+i\eta][-\hbar\omega_2-(\varepsilon_{\mathbf{k},c}-\varepsilon_{\mathbf{k},b})+i\eta]}\\
&\quad+\frac{\bar{f}_{\mathbf{k},a}f_{\mathbf{k},b}\bar{f}_{\mathbf{k},c}}{[\hbar\omega_1-(\varepsilon_{\mathbf{k},a}-\varepsilon_{\mathbf{k},b})+i\eta][-\hbar\omega_2-(\varepsilon_{\mathbf{k},c}-\varepsilon_{\mathbf{k},b})+i\eta]}\\
&\quad-\frac{f_{\mathbf{k},a}\bar{f}_{\mathbf{k},b}f_{\mathbf{k},c}}{[-\hbar\omega_1-(\varepsilon_{\mathbf{k},b}-\varepsilon_{\mathbf{k},a})+i\eta][\hbar\omega_2-(\varepsilon_{\mathbf{k},b}-\varepsilon_{\mathbf{k},c})+i\eta]}.
\end{aligned}
\tag{S17}
$$

It is then straightforward to obtain Eqs. (3-5) from the main text by combining all of the individual pieces and replacing $\eta$ by the electronic broadening $\gamma$.

## S2. COMPUTATIONAL DETAILS FOR THE HBN AND MOS$_2$ CALCULATIONS

Here we present material-specific information regarding the first-principles calculations.

### hBN

The ground state properties of bulk hexagonal boron nitride in the $AA'$ stacking configuration were obtained using norm-conserving, non-relativistic, core-corrected pseudopotentials. We use a plane-wave basis set with an energy cutoff of 110 Ry for the electronic wave functions and 440 Ry for the charge density. For integrations over the first Brillouin zone, a uniform **k**-point mesh of size 12×12×2 was used. The lattice parameters were relaxed prior to the calculation, yielding values of $a$=2.478 Å and $c$=6.453 Å for the in- and out-of-plane lattice constants, respectively. For the electronic excited state properties, we computed the electronic band structure on a 36×36×2 **k**-point mesh for 200 (spin-degenerate) bands. We included the upper two valence and lower two conduction bands in the construction of the two-particle interaction kernel. A rigid band energy shift of 2 eV was added to all independent-particle transition energies in order to account for the underestimation of the band gap in DFT. This value was obtained in previous first-principles calculations on the level of the $GW$ approximation (20). Lastly, for the phonon frequency of the degenerate, Raman-active in-plane optical mode ($E_{2g}$-symmetry), we use the calculated value of 1388.6 cm$^{-1}$.

### MoS$_2$

The ground state properties of monolayer molybdenum disulfide were computed using norm-conserving, fully relativistic pseudopotentials for both sulfur and molybdenum to account for the effects of spin-orbit interactions. We use a plane-wave basis set with an energy cutoff of 90 Ry for the electronic wave functions and 360 Ry for the charge density. For integrations over the first Brillouin zone, a uniform **k**-point mesh of size 12×12×1 was used. The lattice parameter was relaxed prior to the calculation, yielding a value of $a$=3.170 Å for the in-plane lattice constant. To preserve the single-layer nature of the system, we use a vacuum spacing of 16 Å between periodic images of the monolayer. For the electronic excited state properties, we computed the electronic band structure on a 24×24×1 **k**-point mesh with spin-orbit coupling for 400 (in general spin-non-degenerate) bands. We included the upper eight valence and lower eight conduction bands in the construction of the two-particle interaction kernel. A rigid band energy shift of 0.925 eV was added to all independent-particle transition energies in order to account for the underestimation of the band gap in DFT. This value was obtained in previous first-principles calculations on the level of the $GW$ approximation (21). Lastly, for the phonon frequencies of the degenerate, Raman-active in-plane optical mode of $E'$-symmetry and the non-degenerate out-of-plane mode of $A'_1$-symmetry, we use the previously calculated values of 391.7 cm$^{-1}$ and 410.3 cm$^{-1}$ (36).




\clearpage

\end{document}